# Integrated photogrammetric-acoustic technique for qualitative analysis of the performance of acoustic screens in sandy soils

José M. Bravo [1], Fernando Buchón-Moragues [2,*], Javier Redondo [3], Marcelino Ferri [1] and Juan V. Sánchez-Pérez [1]

**Abstract:** In this work we present an integrated photogrammetric-acoustic technique that, together with the construction of a scaled wind tunnel, allows us to experimentally analyze the permeability behavior of a new type of acoustic screens based on a material called sonic crystal. Acoustic screens are devices used to reduce the noise, mostly due to the communication infrastructures, in its transmission phase from the source to the receiver. The main constructive difference between these new screens and the classic ones is that the first ones are formed by arrays of acoustic scatterers while the second ones are formed by continuous walls. This implies that, due to their geometry, screens based on sonic crystals are permeable to wind and water, unlike the classic ones. This fact may allow the use of these new screens in sandy soils, where sand would pass through the screen avoiding the formation of sand dunes that are formed in classic screens and drastically reduce their acoustic performance. In this work, the movement of the sand and the resulting acoustic attenuation in these new screens are analyzed qualitatively comparing the results with those obtained with the classic ones, obtaining interesting results under the acoustic point of view.

## 1. Introduction

Noise can be defined as an unwanted or unpleasant outdoor sound generated by human activity, and is one of the main environmental problems all over the world [1]. This kind of pollution can be controlled in each of the three phases into which its propagation is divided: (i) noise generation at the source; (ii) transmission of noise from source to receiver and (iii) noise reception. The most useful solution to control noise in its phase of transmission from source to receiver is the use of Acoustic Barriers or Acoustic screens (ABs). In Figure 1a one can see an outline of the acoustic performance of these devices. Classical ABs are formed by continuous rigid material with a minimum superficial density of 20 kg/m$^2$ [2] (see an example in Figure 1b). However, these classical ABs present some drawbacks related with aesthetic and communication problems along to their limited technological performance in noise control, as they are not able to discriminate the type of sound and therefore the same AB is used to attenuate very different noises. [3, 4]. In order to obtain ABs with high noise control performance, new advanced materials called sonic crystals are being used, calling the ABs formed by these materials Sonic Crystal Acoustic Screens (SCASs). Sonic crystals can be defined as heterogeneous material consisting of periodic arrays of acoustic scatterers embedded in air [5] (see Figure 1c). A great effort has been made to analyze the physical properties of these new materials and to apply them to the field of environmental acoustic due to their excellent properties in noise control [6, 7, 8], or to analyze their behavior regarding wave propagation in viscoelastic materials instead of in air [9, 10]. One of these interesting properties is their permeability to wind and water as they are discontinuous materials. This permeability can help the placement of ABs in environments that until now were forbidden, as is the case of desert areas, where there is a significant transport of sand due to the existing wind. However, this possibility requires the use of specific techniques to estimate the volume of sand displaced in the presence of SCASs with great precision.

For this purpose, we have used photogrammetry, which can be defined as a set of methods for determining the geometric properties of objects from photographic images. A new photogrammetric technique, called Structured Light System Technique (SLST) [11, 12], has been developed in last years. SLST allows the three-dimensional surface reconstruction of objects with a low cost and with a very high accuracy. SLST is used successfully in a variety of fields, as surgery [13], industry [14] and aeronautics [15], or even in the documentation of cultural heritage [16-19]. Due to its specific characteristics of accuracy and cost, SLST seems the ideal technique to work together with the acoustic part to establish the noise abatement performance of any kind of ABs –classical and SCAS- in sandy soils. The use of integrated photogrammetic-acoustic methods is not new and is increasingly used in recent years in various fields of science and technology. Some examples of this use, without pretending to be exhaustive, have been recently proposed in the characterization of stone building materials, also using laser scanner [20], in the reconstruction of natural and archeological underwater landscapes [21] or topography [22], to estimate the direction of drift and velocity of directional sonobuoys [23], for noise abatement in the design process of aircraft in wind tunnel [24], to develop a method to obtain a dynamic three-dimensional modeling of channel tunnels [25], or finally to predict the collapse and instability of the rocks mass structure in echelon flaws [26].

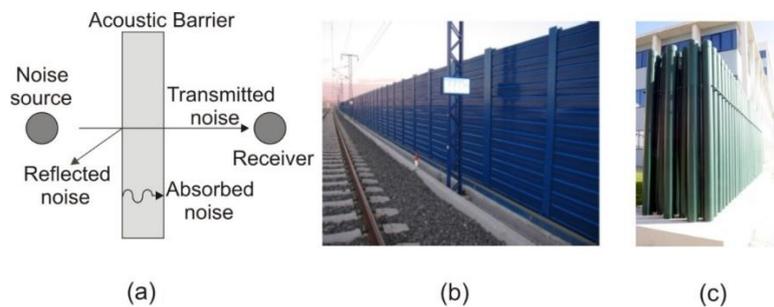

**Figure 1.** (a) Plan view of the acoustic performance of an AB; (b) Image of a classic acoustic barrier, made with a continuous surface, in the surroundings of a railway; (c) Picture of a SCAS made up with rigid cylinders as scatterers, where its transparency can be observed

Following this line, in this paper we present a qualitative study to analyze the variation of the noise attenuation performance of different kinds of SCASs as a function of the sand accumulated in their basis, comparing these results with those obtained with a classical AB. To this end, we have used an integrated photogrammetric-acoustic method consisting of (i) SLST to carefully determine the sand levels accumulated at the base of the different ABs analysed, and (ii) an environmental acoustic measurement system to obtain noise attenuation for each of the proposed cases. Furthermore, using this integrated method we have established a qualitative relationship between the movement of the accumulated sand and the attenuation obtained in all cases.

The article is organized as follows: In section 2 we explain the theoretical basis of SLST. The specific details of the experimental set up, the characteristics of the samples and the stages of the measurement procedure are presented in section 3. In section 4 the results obtained are shown and discussed. Finally, the last section contains the concluding remarks, where the results are summarized.

## 2. Theoretical basis

*2.1. Photogrammetry*

The method used to obtain three-dimensional measurements of the experimental model as a whole has been the automated photogrammetry, also known as photogrammetric correlation.

In order to understand the mathematical principles of the process of measuring from photographic images, it is first necessary to understand that this process of artificial reconstruction is based on the working of the natural binocular vision of living beings (see Figure 2a). In that case, it could be considered that each element of the image captured by each eye could be represented by a line that joins three points: the center of the eye, the point of the retina and the position of the object. The determination of the coordinates of the center of the eye and the point of the retina is not possible in a natural process of vision, but if the eye is replaced by a photographic camera and the image obtained is considered as the point of the retina, it is quite easy to obtain the coordinates of the line that passes through the following three points: the center of the projection of the camera $(X_0, Y_0, Z_0)$, the point of the image $(X_i, Y_i, Z_i)$ and the point of the ground $(X, Y, Z)$, as one also can see in Figure 2a. With these conditions, each projective ray that represents the position of each point of the ground and its counterpart on the image, can be expressed by means of the following equation

$$\left. \begin{matrix} X_i & Y_i & Z_i \\ X_0 & Y_0 & Z_0 \\ X & Y & Z \end{matrix} \right\} \rightarrow \frac{X_i - X_0}{X - X_0} = \frac{Y_i - Y_0}{Y - Y_0} = \frac{Z_i - Z_0}{Z - Z_0} \qquad (1)$$

From equation (1) and considering $Z_i$ as the focal length of the camera used (f), the following expressions are obtained

$$X_i = f \cdot \frac{[m_{11}(X - X_0) + m_{12}(Y - Y_0) + m_{13}(Z - Z_0)]}{[m_{31}(X - X_0) + m_{32}(Y - Y_0) + m_{33}(Z - Z_0)]} \qquad (2)$$

$$Y_i = f \cdot \frac{[m_{21}(X - X_0) + m_{22}(Y - Y_0) + m_{23}(Z - Z_0)]}{[m_{31}(X - X_0) + m_{32}(Y - Y_0) + m_{33}(Z - Z_0)]} \qquad (3)$$

These expressions relate the image coordinates $(X_i, Y_i)$, with the coordinates of the center of projection $(X_0, Y_0, Z_0)$ and the ground coordinates $(X, Y, Z)$. Note that a rotation matrix [M] has been applied in order to place the photographic image in the same position in which it was made on the ground and to force, in this way, the intersection of the projective rays coming from each camera on the same point of the ground (A). Under these conditions, it is considered that the homologous straight line is the one that passes through the central projection points of the right camera $(X_1, Y_1, Z_1)$, point in the right image $(X_d, Y_d, Z_d)$ and the same point of the ground $A(X, Y, Z)$ [27]. An outline of this process can be seen in Figure 2b.

With two images you can artificially reconstruct the natural vision mechanism and obtain the coordinates of the photographed object from the coordinates of the center of each camera and the coordinates of the object on the photographed image. Nowadays, the automatic processes of obtaining ground coordinates by photogrammetric correlation are known as Structure From Motion (SFM). More than two frames are involved in the process of calculating these coordinates. Thus, with a third frame would be included the so-called trinocular epipolar condition [28], where a third projective ray with its origin in the center of projection of a third camera would also intersect at the same point of the ground (A) as the other two projective rays, as can be seen in Figure 2c. In the same way, a fourth frame would include one more condition, called quadrifocal epipolar condition [29-30], and so on (see Figure 2d). In this way, for each camera used two equations would be obtained for each point of the image $(X_i, Y_i)$, generating a system of equations for each point of the image from which its three-dimensional coordinates are to be obtained.

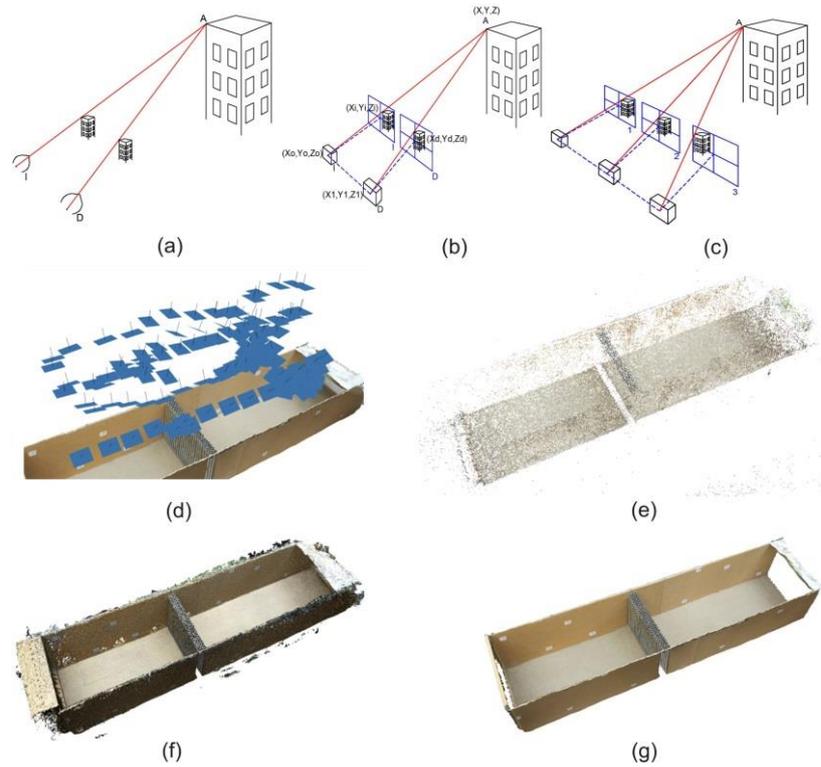

**Figure 2.** (a) Principle of natural stereoscopic vision; (b) Mechanism of artificial stereoscopic vision; (c) Epipolar trinocular condition for automatic correlation of three images; (d) Distribution of the images (in blue) used in one of the measurements of the experimental system. All the images shown meet the corresponding epipolar condition to be placed in the same relative position in which they were taken; (e) Result of a three-dimensional scattered point cloud; (f) Result of a dense three-dimensional point cloud; (g) A three-dimensional photorealistic model obtained after the generation of the triangular mesh and the texturing of its triangles.

The calculation algorithm to obtain the three-dimensional coordinates of an object using the SFM technique consists of the following phases [31]:

(i) Calibration of the photographic camera: The knowledge of the intrinsic parameters of the camera that define the conical projection of the obtained image is crucial to define the geometry of the image. These parameters are: focal distance, position of the main point, radial and tangential distortion, perpendicularity between axes and pixel size and number of horizontal and vertical pixels. The calibration of the camera can be carried out in a previous process in which photographs are taken to already calibrated standards or, on the other hand, if a previous calibration is not carried out, the EXIF metadata file associated with the file of the digital image obtained can be used [32].

(ii) Identification of control points: The success of automatic photogrammetric correlation is the detection of common points in consecutive images. To do this, an initial adjustment is made to all the images to detect a series of common points and ensure that their relative position is the same as at the time of capture [33]. This first adjustment of the frames can be processed with mathematical algorithms as Area Based Matching (ABM) defined as a correlation based on the area, or Featured Based Matching (FBM), which is a correlation based on operators SIFT (Scale Invariant Feature Transform) and SURF (Speeded Up Robust Features)[34].

(iii) Generation of a scattered three-dimensional points cloud: This phase is usually called Bundle Adjustement [35], where the first ground coordinates are obtained from the intersections of the straight lines of each frame of a series of dispersed points of the photographed object. To do this, a block adjustment of the coordinates of the centres of each camera, the control points and those defined for the scattered cloud is carried out. Figure 2e shows the scattered points cloud obtained in the calculation of the three-dimensional model for the case of the thick screen.

(iv) Obtaining the dense point cloud: Starting from the orientation elements obtained in the adjustment of the disperse cloud to improve the coordinates of the projection centres, the photogrammetric correlation is carried out pixel by pixel. For this purpose, algorithms of the type Path Based Multi-View Stereo (PVMS) are used, which is based on an adjustment by least squares with geometric constraints [36], Multi Image Correspondance par Méthodes Automátiques de Corrélation (MIC MAC), which allows the correlation of points from pyramidal images [27], or Semi Global Matching (SGM), which performs calculations by pairs of images, as we show in Figure 2f [37].

(v) Reconstruction of the surface and texturing: Finally, and starting from the dense cloud of points obtained, a three-dimensional mesh is made using triangles [38] which will be provided with the tonality that defines the photographic image for that area [39], obtaining realistic three-dimensional models (see an example in Figure 2g).

## 2.2. Acoustics

The behavior of an acoustic system is characterized by its transfer function, H(f), that can be obtained as the Fourier transform of the impulse response, h(t). If the impulse response is known, when the system is excited by an arbitrary input, the output signal can be calculated as follows:

$$o(t) = i(t) \otimes h(t) = \int_{-\infty}^{\infty} i(\tau) \cdot h(t - \tau) \cdot d\tau \qquad (4)$$

where o(t) is the time domain output, i(t) is the time domain input and the symbol $\otimes$ stands for convolution. Equation (2) can be translated to the frequency domain as follows:

$$O(f) = I(f) \cdot H(f) \qquad (5)$$

where O(f) is the output of the complex frequency domain and I(f) is the input of the complex frequency domain. Assuming that the system is linear and invariable in time, the impulse response can be measured using a Sine Sweep signal. This technique has been employed since long time for audio and acoustics measurements, but in last two decades it became more popular within the scientific community, thanks to the computational capabilities of modern computers and the additional possibility of measuring simultaneously the distortion [40].

In our case, a sine sweep starting at 500Hz and finishing at 20kHz with a total length of 10s has been emitted by the loudspeakers in all the possible configurations (with and without the different ABs and after the movement of sand under wind conditions with every AB placed), and then recorded at the receiver position. The mathematical definition of the test signal is as follows:

$$x(t) = sin\left[\frac{2\pi f_1}{\ln\frac{f_2}{f_1}} \cdot \left(e^{\frac{t}{T}\ln\frac{f_2}{f_1}} - 1\right)\right] \qquad (6)$$

where $f_1$ and $f_2$ are the lowest and the highest considered frequencies respectively.

In order to obtain the impulse response in all the situations under analysis, the recorded signals have to be convolved with a filter that returns the impulse-response. This is equivalent to a deconvolution. The filter is just a time reversal version of the excitation signal. We have also taken into account the fact that the test signal has not a white (flat) spectrum: given that the instantaneous frequency sweeps slowly at low frequencies and much faster at high frequencies, the resulting spectrum is pink (falling down by -3dB/octave). To compensate this coloration of the signal a filter has been used, applying a proper amplitude envelope to the reversed sweep signal so that its amplitude is now increasing by +3dB/octave. Figure 3(b) illustrates the impulse-response spectra obtained in two representative cases.

The different impulse-responses spectra have to be translated to the frequency space by a Fourier transform, averaged in frequency bands and then the Insertion loss can be calculated. Namely,

$$IL = 20 \cdot \log_{10}\left(\frac{p(without\ barrier)}{p(with\ barrier)}\right) \quad (7)$$

where p stands for the acoustic pressure.

The insertion loss (IL) parameter has been used for the acoustic assessment. IL can be defined as the difference between the sound pressure level ($L_P$) created by noise source, recorded at the same point with and without the AB

$$IL = L_P(without\ AB) - L_P(with\ AB) \quad (8)$$

Figure 3a illustrates the positioning of the acoustic sources and the receiver near the considered AB.

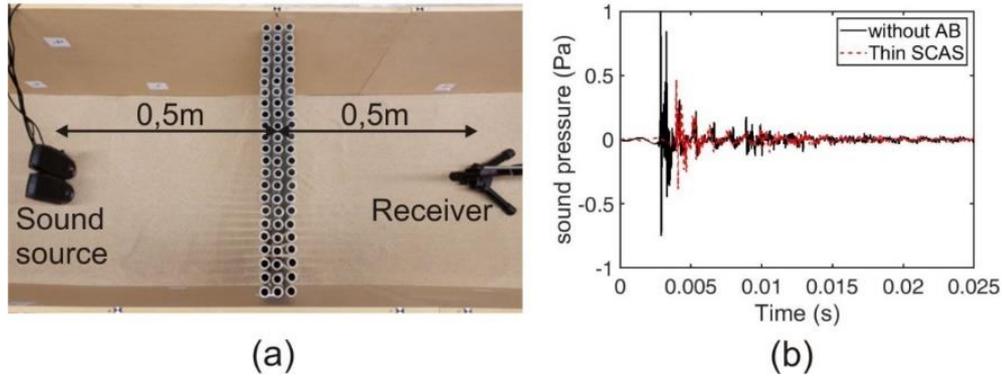

**Figure 3.** (a) Plan view of the experimental set up (loudspeakers and condenser microphone) and the position of the different AB; b) Examples of the impulse-response spectra for two cases: without AB (black solid line) and with thin SCAS (red dashed line).

## 3. Materials and Methods

As mentioned in the introduction, the main objective of the work is to measure, in a sandy environment, the acoustic attenuation obtained by different ABs (classics and SCASs) taking into account the sand accumulated at the base of each AB by airflow. The measurement of the amount of sand accumulated is done with SLST to obtain the most accurate values. In order to carry out this work, it has been necessary to develop a complex prototype on a scale that allows, under controlled conditions, the dragging of the sand by means of airflows, the subsequent measurements of its accumulation at the bottom part of the different ABs designed, and finally the acoustic evaluation of each prototype tested under the different conditions of sand accumulation.

This section presents the different materials used for the development of the designed prototype, divided into the following items: (i) Wind tunnel; (ii) Selected sand; (iii) Air fan and diffuser; (iv) AB and SCASs; (v) Material used for the acoustic measurements; (vi) Material used to measure the sand distribution around the barriers.

### 3.1. Scaled wind tunnel

The scaled wind tunnel is the core of the experiment, and its function is to contain the sand when it is moved by the air fan, orienting it towards the ABs. It is designed in the form of a quadrangular prism resting on one of its lateral faces, and is uncovered both by the upper lateral face and by the two bases, as can be seen in Figures 4a- 4b. It is made of medium density fibreboard (MDF).

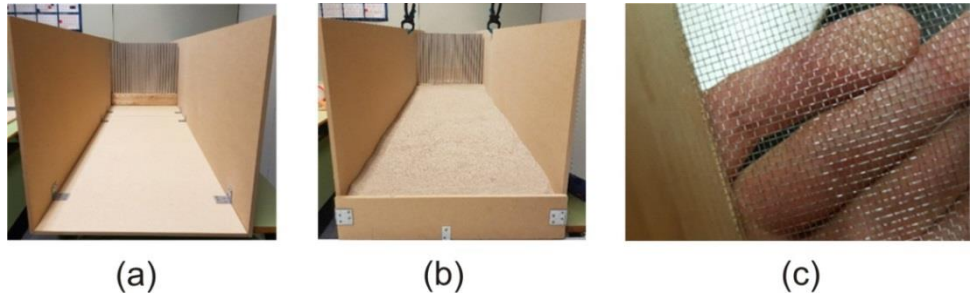

**Figure 4.** View of the scaled wind tunnel with a SCAS prototype (a) without sand and (b) with the sand reservoir used (b); c) Sieve used for sand homogenization.

The scaled wind tunnel is formed by (i) 6 boards of 122x52x1cm³ corresponding to three of the lateral surfaces of the square prism and (ii) 2 boards of 10x52x1cm³ placed at the beginning and at the end of the tunnel (prism bases) to allow a storage of sand 10cm deep. With these dimensions, the final size of the scaled wind tunnel constructed is 244x52x52cm³

The panels are assembled and sealed with silicone to prevent the loss of sand through the junctions. At the end of the tunnel there is a bag that will serve to store the sand expelled by the airflow created by the fan.

*3.2. Sand*

The scaled wind tunnel has been designed with a reservoir to accumulate a 10cm high amount of sand to cover its base. Taking into account the dimensions of this base (244x52cm²) and the average density of beach sand used (1600kg/m³), a total amount of 200kg of sand was needed to fill the reservoir.

The use of beach sand implies the existence of sticks, algae, shells and other unwanted objects. To eliminate these impurities and to homogenize the type of grain, the sand used in the experiment has been sieved. The sieve used has a square mesh opening of 1 mm, with a Tyler mesh number (mesh/line) of 16 (see details in Figure 4c).

*3.3. Fan and air difusser*

A set of preliminary tests has been carried out to analyze the homogenization in the sand distribution along the scaled wind tunnel due to the airflow created by the fan used. An example of the results obtained can be seen in Figure 5a, where a trend of sand accumulation is observed on the right side of the wind tunnel. This phenomenon is caused by the airflow created by the movement of the blades, which in its encounter with the walls of the prototype creates a small turbulence on one side of the tunnel. To prevent this effect and to try that the prototype works with a more laminar flow similar to the one we can find in nature, an air diffuser has been designed and built specifically to channel and laminate the airflow expelled by the fan.

The designed diffuser has two parts, and is prepared to fit into the tunnel inlet with a surface covering the diameter of the fan to channel most of the impelled airflow. The diffuser has been designed to direct the airflow, and is made up of 12 movable sheets of wood 0.5cm thick with a height of 40cm high and 4cm separation between them. Figures 5b-5c show different views of the final aspect of the constructed diffuser and its location in the prototype.

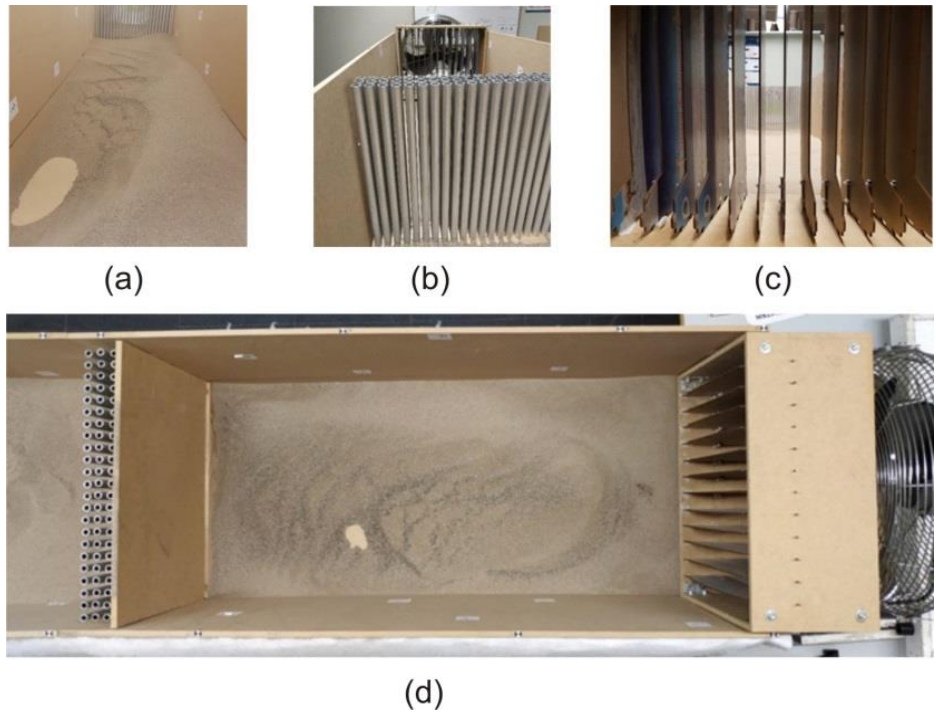

**Figure 5.** (a) Accumulation of sand on the right side of the tunnel caused by the turbulences of the fan blades before the insertion of the diffuser; (b) Front view of the diffuser in the final assembly; (c) Back view of the diffuser placed in the tunnel; (d) Plant view of the scaled wind tunnel

*3.4. Acoustic screens tested: Classical and Sonic Crystal acoustic screens*

Three ABs, one classic and two based on the sonic crystals technology, have been designed and built to carry out this work.

The classical AB, which by definition is a continuous wall, consists of an MDF panel of 37.5x52x1cm³ and is placed in such a way that the transversal section of the scaled wind tunnel is completely covered. This kind of ABs is usually designed to attenuate noise in a wide range of frequencies due to its geometric characteristics, as it completely covers the line of sight between the sound source and the receiver avoiding the transmission of noise.

On the other hand and as previously explained in the introduction section, sonic crystals can be defined as a heterogeneous material formed by arrays of acoustic scatterers embedded in air. These materials present several properties related to the control of acoustic waves [41-43], but the main one related to noise attenuation and used in the design of ABs is the existence of bandgaps in the frequency domain. Bandgaps are frequencies ranges where acoustic waves cannot be transmitted through sonic crystals, and their existence is due to Bragg's law [44, 45]. Their size and position in the frequency domain depend on the geometric parameters of the sonic crystals, such as (i) the kind of crystalline array in which the scatterers are arranged; (ii) the distance between them (generally known as the lattice constant) or (iii) the surface area of the scatterers per surface unit of the crystalline lattice (called filling factor). This fact implies that, while classical ABs work attenuating a wide part of the frequencies spectrum, SCAS act as a filter, attenuating a specific and limited frequencies range. The different way of working of both kind of ABs (classical and SCAS) makes it very difficult to carry out a quantitative study on how the movement of sand in the bases of the ABs influences the acoustic attenuation achieved, since this attenuation is not comparable in any case.

The design of the SCASs used in this work is based on those most used in the field of noise attenuation, which present a great symmetry and simplicity in their design [46]: hollow cylindrical aluminium scatterers arranged in a square array. In order to have a greater variety of

SCAS to be tested, two different diameters of the scatterers have been considered, calling them thick SCAS (larger diameter) and thin SCAS (smaller diameter). The design of these SCASs has been made to real size and then adjusted to the dimensions of the wind tunnel, applying a scaling coefficient. In this study, the 1kHz Bragg frequency has been considered as the center of the bandgap, due to it is the most important band in the normalized traffic noise spectrum [47]. With this condition, the lattice constant adopted is 17cm and the diameter of the cylinders is 13,6cm, for a filling factor of 0.5, appropriated for obtaining an acceptable bandgap size [45].

From the reference SCAS to real size, the parameters have been adjusted for a scaling factor of 1:8. Taking into account that the Bragg frequency will also be transferred according to the scale factor, the Bragg attenuation frequency for these scaled SCASs will be in the 8 kHz band. By setting a filling factor of 0.25 (thin SCAS) and 0.45 (thick SCAS) to analyze different cases, the resulting design parameters of both scaled SCAS are shown in Table 1:

**Table 1.** Design parameters of thin and thick barriers based on sonic crystals at scaling factor 1:8.

| Name | Scaling factor | Bragg Frequency (Hz) | Lattice constant (cm) | Tube diameter (cm) |
|---|---|---|---|---|
| Thin SCAS | 1:8 | 8000 | 2,125 | 1.2 |
| Thick SCAS | 1:8 | 8000 | 2,125 | 1.6 |

Two commercial diameters of aluminum cylinders, 16mm and 12mm, have been used for the design of thick and thin SCASs respectively. The SCASs prototypes have been built to perfectly fit the cross section of the tunnel. To take into account the 10cm deep sand reservoir, the cylinders are assembled on a wooden base of 52x10x10cm³. In Figure 6 you can see different geometric details of both scaled SCAS.

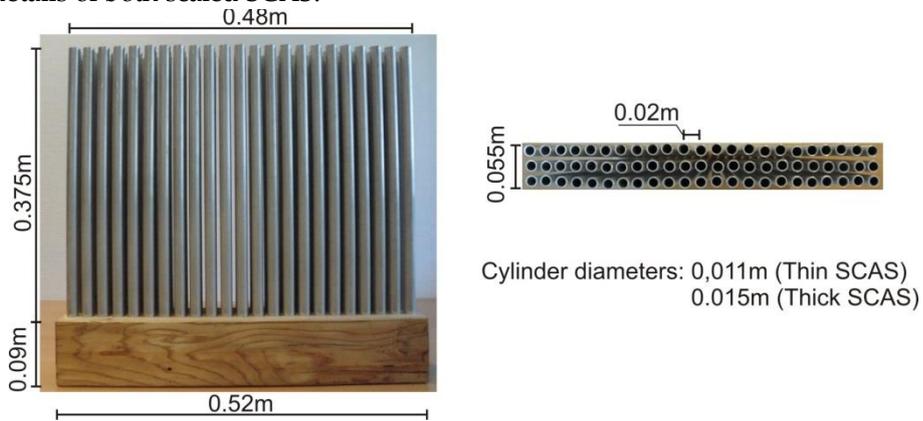

**Figure 6.** Front and plan view of the scaled SCASs.

*3.5. Acoustic measurements equipment*

The materials used for the evaluation of the IL of the different ABs basically consist of a sound emitter, a sound card that allows the audio signal to be digitized, and software that allows both the creation of the emitted signal (sine sweep) and the recording of the audio tracks for further analysis. The details of this material can be seen in table 2.

**Table 2.** Material used in acoustic measurements (IL).

| Name | Image | Features |
|---|---|---|
| Loudspeakers | 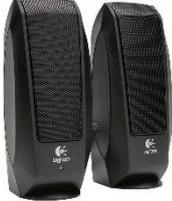 | Total Watt (RMS): 2.2W |
| Condenser microphone | 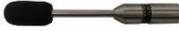 | Prepolarized<br>Nominal Sensitivity 6mV/Pa<br>Frequency Response 10-20.000Hz<br>Class 1 |
| Soundcard | 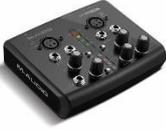 | XLR Inputs<br>Signal to Noise Ratio: 97dB<br>THD+N: 0.005%<br>Freq. Response: ±0.35dB |
| Software Cool Edit 2000 | 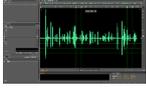 | Record signal<br>Manipulating signal<br>Mix signal |

*3.6. Photogrammetric measurements equipment*

The materials used to carry out the three-dimensional models and the corresponding measurements of the longitudinal sections of the topographic profiles adopted by the sand in each case of the ABs tested have been (i) the photogrammetric correlation software Photoscan from Agisoft (ii) TRIMBLE three-dimensional software for the treatment and management of point clouds and (iii) Canon EOS1DX MKIII camera, with a resolution of 21.1Mpx, 35mm format, CMOS sensor and a 50mm lens for image acquisition.

## 4. Results

In this section we analyze the results obtained in the experiments carried out with the different ABs prototypes designed on a scale of 1:8, and then extrapolate these results to a real scale of 1:1.

The experimental protocol has been developed as follows: First, the sand reservoir of the scaled wind tunnel was filled with 10cm of clean and sieved sand. The sand has been flattened as much as possible to ensure that it is as horizontal as possible at all points in the wind tunnel. With this configuration, and without ABs located in the tunnel, the sound pressure was measured with the equipment indicated in section 3.5. Next, and for each of the three considered ABs, photogrammetric and sound pressure measurements were made before and after connecting the scaled wind tunnel fan. The photogrammetric measurements provide us the sand permeability behaviour of the three ABs. In the same way, the acoustic measurements allow us to quantify the shift of the acoustic attenuation capability due to the variation of the sand relief for each of the ABs tested. In the different experiments the fan worked for three hours in a row, the time needed to exhaust the sand tank.

The results obtained are divided into two categories, which are undoubtedly interrelated and not independent, but which will be explained separately for a better understanding of the data obtained and a simpler interpretation of the results. First, the permeability of the different ABs and their effect on sand movement will be evaluated. This aspect is particularly interesting if SCASs are to be installed in desert lands, for example in infrastructures such as the AVE Medina-La Meca in Saudi Arabia (Haramain High Speed Rail o HHR). In this type of terrain, the placement of classical ABs, formed by walls, can result in significant sand accumulations that endanger the transport infrastructures for which they are designed. However, SCASs would allow sand to pass through them, thus avoiding sand accumulations and protecting the infrastructure. This first analysis will allow us to quantify these accumulations for both types of ABs, the classics and the SCASs, under controlled conditions, as well as to make a first evaluation of the goodness of the SCASs with respect to the classical ABs in this sense. On the other hand, it is necessary to analyze the effect of sand accumulation on the acoustic efficiency of the different ABs considered. For this purpose, the acoustic attenuation of each AB will be measured in each of the cases studied: before and after the movement of sand due to the airflow created in the scaled wind tunnel. However, as we have indicated in section 3.4, the quantitative comparison of both types of ABs is very difficult due to the different way they work acoustically. Therefore, here we can only compare qualitatively the effect of sand on the attenuation properties of the AB classics and SCAS separately. But this first study on the relationship between the movement of sand and the change in acoustic attenuation properties of SCAS versus classical ABs may be very interesting for engineers to select the type of ABs to be installed in communication infrastructures depending on the type of ground.

*4.1. Photogrametric Results*

In this section we have analyzed the sand permeability of the different AB prototypes considered. To this end, we have compared the initial flat surface of the sand with its relief for each AB, after being subjected to the airflow of the wind tunnel, on both sides of each AB.

To quantitatively estimate the cubicage of sand moved in this comparison, we have considered three longitudinal sections of the scaled wind tunnel (called sections 0.00, 0.26 and 0.52), as can be seen on the left side of Figure 7a. Only three longitudinal profiles have been taken into account for the calculation of this cubicage since the displacement of the sand is regular, without large level differences in the wind tunnel cross-sections (in the direction of the ABs), due to the use of the air diffuser.

On the other hand, Figures 7b-7d show, for each of the ABs considered, (i) the three longitudinal sections with the starting (black line) and modified (red line) sand profiles; (ii) section 0.26 indicating the clearing (pink color) and embankment (blue color) of the sand and finally (iii) a detail of this last section. In Figure 7 you can also see qualitatively the variation in the sand distribution in front and behind each of the tested ABs, providing a clear picture of their permeability behaviour.

The cubicage of sand moved due to airflow in the scaled wind tunnel for an each AB case has been calculated following the standard method of cubication in linear works, i.e. calculating the volume of sand moved in each clearing zone or embankment between each of the two sections considered. A diagram of how the measurements have been made can be found on the right side of Figure 7a.

The results of the cubicage performed are shown in table 4. The analysis of this table leads to results that can be considered as expected. Thus, first of all it is observed that in the classic AB the main sand embankment takes place in its front part, that is, in the side exposed to the wind. The fact that classic ABs are formed by continuous walls, without openings, implies that the sand driven by the airflow accumulates in its front face, being practically insignificant in the back.

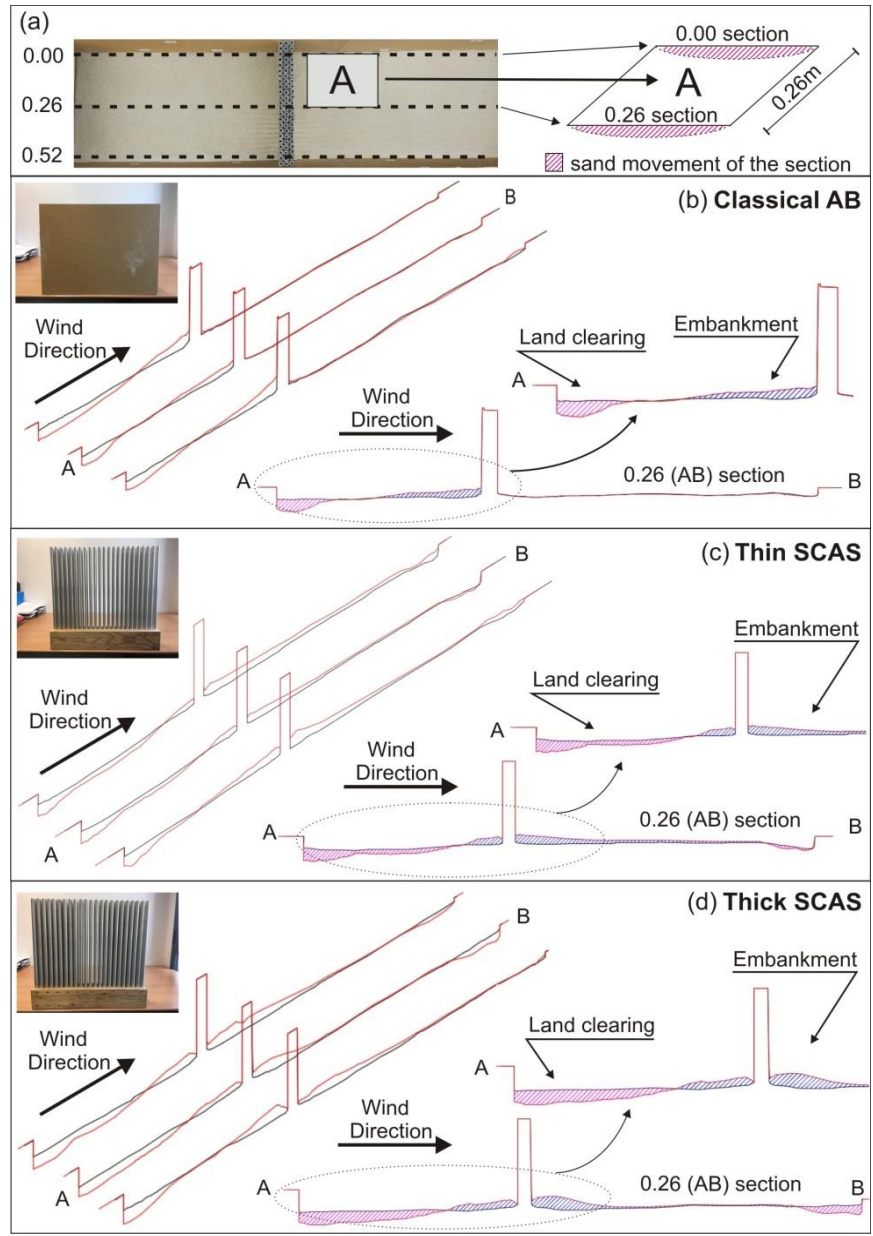

Figure 7. (a) longitudinal sections used to evaluate the clearing or embankment of sand on both sides of each AB; (b)-(d) For each AB: the three longitudinal sections considered shown in perspective and the 0.26 profile image with enlargement of the area close to the AB.

In the case of SCAS, the results are more interesting. First, it can be seen that the filling factor (index indicating the amount of mass of scatterers in the crystal) significantly affects the passage of sand. Thus, it can be observed that in the case of the thick SCAS there is a 28.5% increase in the embankment of sand in its front part compared to what happens in thin SCAS, being the filling factor of the first almost doubles that of the second (0.45 vs. 0.25). On the other hand, in the back part of the thick SCAS the sand embankment produced is also larger than in the thin SCAS (11,6%). With these results, it can be concluded that increasing the filling factor in a SCAS will increase its sand accumulation capability and therefore its behavior will be more and more similar to that of classic ABs, increasing the "dune effect" that appears with its maximum value in the front part of these last ones.

**Table 4.** Cubication of the sand displaced by the effect of the airflow in the scaled wind tunnel for each of the ABs tested both on a full scale (1:1) and on a scale of 1:8.

| | | VOLUMES (m³) | | | |
| --- | --- | --- | --- | --- | --- |
| | | FRONT PART OF THE AB | | BACK PART OF THE AB | |
| | PROFILES | CLEARING | ENBANKMENT | CLEARING | ENBANKMENT |
| CLASSIC AB | 0,00-0,26 | 0,0030 | 0,0031 | 0,0008 | 0,0004 |
| | 0,26-0,52 | 0,0038 | 0,0027 | 0,0000 | 0,0004 |
| | Scale 1:8 | 0,0068 | 0,0057 | 0,0008 | 0,0008 |
| | Scale 1:1 | 0,0136 | 0,0114 | 0,0017 | 0,0016 |
| THIN SCAS | 0,00-0,26 | 0,0035 | 0,0018 | 0,0003 | 0,0067 |
| | 0,26-0,52 | 0,0088 | 0,0016 | 0,0010 | 0,0033 |
| | Scale 1:8 | 0,0123 | 0,0034 | 0,0012 | 0,0100 |
| | Scale 1:1 | 0,0246 | 0,0068 | 0,0025 | 0,0200 |
| THICK SCAS | 0,00-0,26 | 0,0051 | 0,0026 | 0,0010 | 0,0058 |
| | 0,26-0,52 | 0,0065 | 0,0022 | 0,0015 | 0,0055 |
| | Scale 1:8 | 0,0116 | 0,0047 | 0,0024 | 0,0113 |
| | Scale 1:1 | 0,0231 | 0,095 | 0,0049 | 0,0226 |

*4.2. Acoustic Results*

In this section, we have analyzed how the change in the relief of the sand due to airflow influences the acoustic attenuation capability of the tested ABs.

To do that, we have calculated the IL index for every AB as a function of the frequency before and after the change in the sand relief due to the airflow. The results can be seen in Figure 8, where the comparison of IL for each AB before (red line) and after (black line) airflow is presented, both on the scale of the experiment (1:8) and on a real scale (1:1).

In the case of the classical AB (Figure 8a), we can observe that a generalized decrease of IL is produced at the frequencies range considered after the airflow, when an improvement of the sand accumulation on the front side of the AB is produced, except in the 350Hz-500Hz range where a slight improvement of the IL before the airflow appears.

To analyze the acoustic attenuation behavior of SCASs, it is worth remember that they have been geometrically designed in such a way that the bandgap must appear around 1kHz, which in the scale model is transferred to an 8kHz bandgap. This means that the only attenuation we must take into account is the one that appears in this band, because it is the one that is directly related to the behavior of the sonic crystal. No other possible attenuations that appears in the spectra and that is not directly due to the action of the sonic crystals has been considered in this work. In the IL spectra of both SCASs, shown at Figures 8b- 8c, an interesting effect in their attenuation properties appears: the size of the bandgap increases after the airflow when the sand is accumulated on the basis of the SCAS, regardless of the value of filling factor. Thus, this effect can be considered as robust although is especially evident for the higher filling fraction, due to the thick SCAS achieves a very noticeable increase in IL around the bandgap frequency of 8kHz (1kHz at real scale) of up to 5dB, while the thin SCAS prototype presents an increase of only 2-3dB around this bandgap frequency.

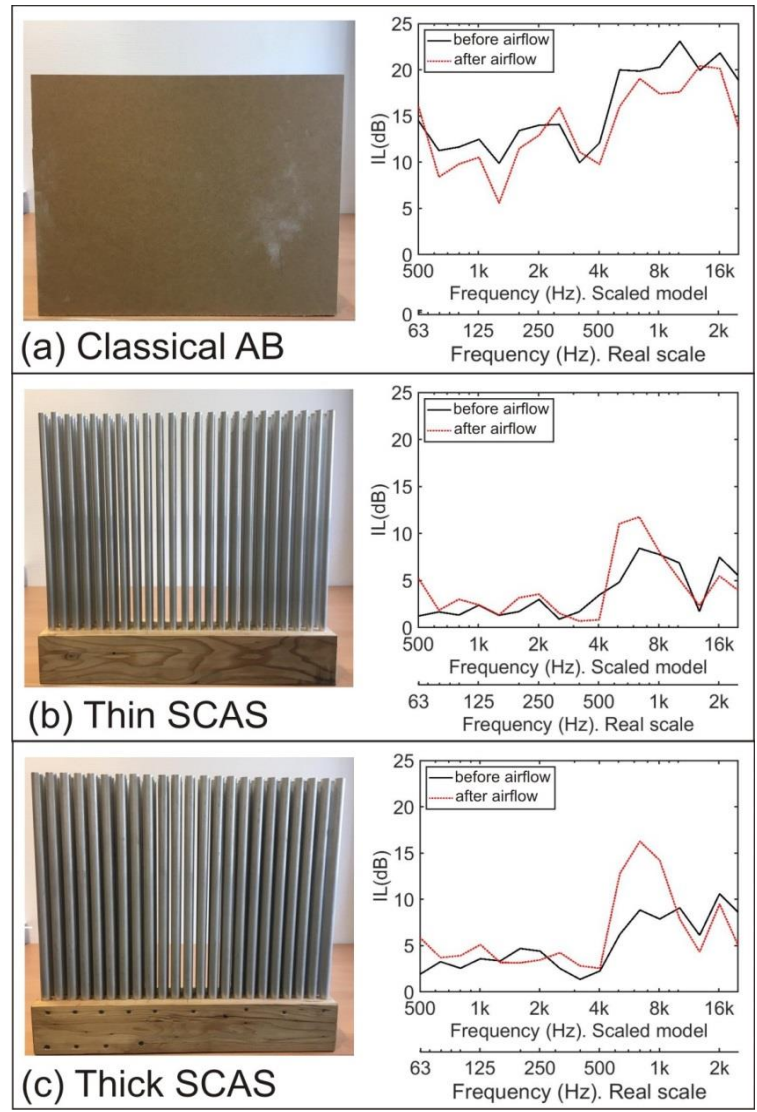

Figure 8. For every AB tested: At left hand image of screen. At right hand values of IL in relation with frequency, before (black continuous line) and after (red dashed line) airflow

## 5. Conclusions

In this work we present a mixed photogrammetric-acoustic technique to carefully analyze, together with the design and construction of a scale wind tunnel, the attenuation behavior and the permeability to sand of a new type of acoustic barriers based on periodic arrays of scatterers, usually called sonic crystals acoustic screens. In contrast to the non-permeability properties of classic barriers, the screens based on sonic crystal allow permeability to wind and water, but until now no one has analyzed the sand permeability and its influence on sound attenuation. This situation is especially interesting if we think in infrastructures located in desert soils.

The development of the work has been complex, because the quantitative comparison of the results in both types of ABs is very difficult due to the different way they work acoustically. Therefore, for this first study on the relationship between sand movement and the change in acoustic attenuation properties of SCAS versus classical ABs, we preferred to focus on a qualitative analysis, which may already be of interest to engineers when selecting the type of ABs to be installed in communications infrastructure depending on the type of soil.

Thus, we have experimentally verified that (i) the permeability of these screens to sand is high, allowing its passage and avoiding the accumulation of sand at the base of these devices forming dunes, as happens in the case of classic screens; (ii) in this new kind of barriers, the attenuation performance increases as a function of the filling factor when the sand accumulates at the base of the screens unlike what happens with traditional screens, which decreases when the sand is accumulated. Although the physical reasons for this behavior are not yet known, the fact reported seems very interesting under an engineering point of view.

However, more funds are needed to build a battery of sonic crystal acoustic screen prototypes in order to analyze their permeability and acoustic performance in the scaled wind tunnel. At the same time, it seems necessary to improve the scaled wind tunnel in order to obtain better performance, for example, to increase the sand storage capability and analyze the distribution of sand in a higher period of time. In addition, it will be necessary to carry out more studies in free field conditions and to develop powerful simulators to understand and use the mechanisms that generate the increase in attenuation performance due to the accumulation of sand at the base of this kind of screens.